\documentclass[11pt,letterpaper]{article}
\pdfoutput=1

\usepackage{jcappub}

\usepackage{amssymb}
\usepackage{graphicx}
\usepackage{amsmath}
\usepackage{wasysym}
\usepackage{verbatim}

\font\FermiSmallfont=cmssq8 scaled 1200

\def\UMDppthead#1#2#3{
\null 
\begin{center}\vskip -1.0truein{\hbox to 7.5truein {
\hfill
\vbox to 1in {\vfill \FermiSmallfont
              \hbox{#1}
              \hbox{#2}
              \hbox{#3}
              \vfill}
}}\vskip-0.0truein\end{center}}

\begin{document}


\title{Models of the Contribution of Blazars to the Anisotropy of the Extragalactic Diffuse Gamma-ray Background}

\author[a,b]{J.\ Patrick Harding} \emailAdd{hard0923@umd.edu}
\author[b]{Kevork N.\ Abazajian} \emailAdd{kevork@uci.edu}
\affiliation[a]{Maryland Center for Fundamental Physics \& Joint
  Space-Science Institute, Department of Physics, University of
  Maryland, College Park, Maryland 20742 USA}
\affiliation[b]{Center for Cosmology, Department of Physics \& Astronomy,
  University of California, Irvine,\\ Irvine, California, 92697 USA}

\date{August 1, 2012}

\abstract{
We study the relation between the measured anisotropies in the  extragalactic diffuse gamma-ray background (DGRB) and the DGRB spectral intensity, and their potential origin from the unresolved blazar population.  
Using a physical-evolution model for blazars with a luminosity dependent density evolution (LDDE) and an observationally-determined luminosity-dependent blazar spectral energy distribution (SED), we find that blazars can account for the observed anisotropy of the DGRB consistent with their observed source-count distribution, but are in turn constrained in contributing significantly to the observed DGRB intensity. 
For the best-fit LDDE model accounting for the DGRB anisotropy and source-count distribution, blazars only contribute $5.7^{+2.1}_{-1.0}\%$ (68\% CL) of the DGRB intensity above 1 GeV. 
Requiring a higher fraction of the DGRB intensity contribution by blazars overproduces the DGRB anisotropy, and therefore blazars in the LDDE+SED-sequence model cannot simultaneously account for the DGRB intensity as well as anisotropy.
We discuss the limitations of LDDE models.  
However, these models do not require the many unjustified and observationally-inconsistent simplifying assumptions---including a single power law for all blazar spectra  and a simple broken power-law model for their source-count distribution---that are present in much previous work.
}


\maketitle

\section{Introduction}

The launch of the Large Area Telescope (LAT) aboard the Fermi Gamma Ray Space Telescope~\cite{collaboration:2011bg} has ushered in a new era of an accurate understanding of the extragalactic gamma-ray sky.
Blazars are the most numerous extragalactic gamma-ray sources seen by Fermi-LAT, yet the contribution of blazars to the extragalactic gamma-ray sky remains an open question. 
Below the Fermi-LAT point-source resolution threshold, the inferred total gamma-ray flux in the diffuse component due to blazars is highly model dependent, and is also every dependent on the nature of the blazar population with the spectrally-dependent ability of Fermi-LAT to resolve point sources.

Blazars are comprised of two classes of active galactic nuclei (AGN): flat-spectrum radio quasars (FSRQs) and BL Lacertae objects (BL Lacs). 
Both of these sources have highly variable luminosity and high luminosity~\cite{collaboration:2011bg}, with a bolometric luminosity which is dominated by their gamma-ray luminosity~\cite{Fossati:1997vu,Fossati:1998zn,Donato:2001ge}, so they are believed to come from a single source class, blazars. 
A blazar occurs when an AGN is observed down the relativistic jet rather than the usual observation of the accretion disk surrounding the AGN. 
Unlike the near-isotropic flux coming from the accretion disk, the emission from the jet is expected to be relativistically beamed~\cite{Blandford:1979za,Dermer95}.

Several models of the blazar population have been proposed in the literature~\cite{Padovani93,Stecker93,Salamon94,Chiang95,Stecker96,Chiang98,Mucke00,Giommi05,Narumoto06,Dermer07,Pavlidou08,Bhattacharya09,Collaboration:2010gqa,Abazajian:2010pc,Stecker:2010di,Malyshev:2011zi}. 
Physical-evolution models are formed by considering the blazar evolution and their constituent spectral dependence. 
A pure luminosity evolution model of blazars considers only the evolution of blazar luminosity with redshift~\cite{Chiang95,Chiang98,Dermer07,Bhattacharya09}. 
Luminosity-dependent density evolution (LDDE) models also allow blazar peak luminosity to vary with redshift~\cite{Ueda:2003yx,Hasinger05}. 
These models are based on the observed properties of X-ray AGN and relate the gamma-ray properties of the blazar model to these X-ray AGN measurements. 
Some blazar models, such as those from Ref.~\cite{Collaboration:2010gqa} (hereafter FB10) and Ref.~\cite{Cuoco:2012yf} (hereafter CKS), assume that the measured Fermi-LAT blazar spectral index distribution is largely independent of blazar flux and that the measured Fermi-LAT source-count distribution is a power law that continues unaltered down to zero flux.
In addition, the spectral properties of blazars are often assumed to be a simple power-law or a distribution of power-laws, but more detailed frequency-dependent models have been used as well~\cite{Giommi05,Inoue:2008pk,Abazajian:2010pc}. 
These frequency-dependent blazar spectra provide a more realistic fit to measured blazar properties, seen in the spectral energy distribution (SED) sequence of Ref.~\cite{Fossati:1997vu,Fossati:1998zn,Donato:2001ge}.

In this work, we study the LDDE with a SED sequence blazar model like that in Ref.~\cite{Abazajian:2010pc} (hereafter ABH). 
The use of an LDDE blazar evolution model incorporates explicitly the blazar evolution in both redshift and luminosity. 
The blazar SED sequence gives the energy-dependence of the blazar luminosity and accounts for differing spectra for blazars of different luminosity. 
Only such a physically-motivated model of the blazar evolution and spectrum can accurately predict the blazar flux below the Fermi-LAT threshold.
In fact, the measured extragalactic diffuse  gamma-ray background (DGRB) by the Fermi-LAT Collaboration~\cite{Abdo:2010nz} was predicted in shape and amplitude by Inoue \& Totani~\cite{Inoue:2008pk} using the LDDE plus SED sequence blazar evolution model~\cite{Inoue10a}.
\footnote{Note we use the term  extragalactic diffuse gamma-ray background (DGRB) instead of the {\it isotropic} diffuse extragalactic gamma-ray background because the DGRB can have, and has been measured to have, anisotropies.}

In particular, the properties of the blazar evolution and spectrum strongly affect the blazar contribution to the DGRB. 
The DGRB is derived by measuring the full gamma-ray spectrum, then removing resolved point-sources and a model for the Galactic diffuse flux. 
The remaining background, the DGRB, is made up of unresolved extragalactic point-sources, such as blazars, and the isotropic component of unmodeled Galactic sources, such as millisecond pulsars or, potentially, annihilating dark matter. 
The most accurate measurements of the DGRB have been made with the Fermi-LAT~\cite{Abdo:2010nz,Collaboration:2010gqa,Xia:2011ax,Ackermann:2012uf}. 

The DGRB spectrum measured by the Fermi-LAT has an intensity $I(>100{\rm\ MeV})=1.03\times10^{-5}\rm\ ph\ cm^{-2}\ s^{-1}\ sr^{-1}$ and follows a power-law in energy with index 2.41~\cite{Abdo:2010nz,Collaboration:2010gqa}. 
This was determined by observing energies from 200 MeV to 100 GeV at angles above $10^\circ$ in Galactic latitude ($\lvert b\rvert>10^\circ$). 
The DGRB flux accounts for roughly 25\% of the total observed Fermi-LAT emission. 

Recently, Ref.~\cite{Ackermann:2012uf} (hereafter FA12) measured the anisotropy of the DGRB. 
Taking the angular power spectrum of the DGRB in energies from 1 GeV to 50 GeV, they measure that the angular power above multipole moment $\ell>155$ is nearly constant in $\ell$. This corresponds to angular scales of $\sim 2^\circ$. The angular power observed in these high multipoles is significantly less than the noise due to finite measurement statistics expected for the Fermi-LAT. Therefore the Poisson term in the angular power spectrum is expected to come from unresolved sources rather than from measurement noise.
For the {\sc DATA:CLEANED} sample, the anisotropy is given in our Table~\ref{CPtable}. 
For many source classes, such as blazars, the DGRB anisotropy limits the total intensity of unresolved sources, which in turn constrains the contribution of that source class to the DGRB intensity. Here, we consider the contribution of unresolved blazars to the DGRB anisotropy and the corresponding contribution of such blazars to the DGRB intensity.

Note that FA12 measured the anisotropy considering {\it all} the diffuse Fermi-LAT flux, including the Galactic diffuse which was removed in the DGRB measurement of Ref.~\cite{Abdo:2010nz}. 
Therefore, the intensity measured by FA12 differs from the DGRB spectral intensity in Ref.~\cite{Abdo:2010nz}. 
However, in stating the dimensionful anisotropy $C_P$, the Galactic diffuse emission is expected to only contribute at low $\ell$, so the value of $C_P$ effectively neglects the Galactic diffuse emission and is consistent with the DGRB. 
Therefore, in this work, we compare calculations of the {\it dimensionful} $C_P$ rather than the {\it dimensionless} $C_P/\langle I\rangle^2$. 
Also, because the anisotropy measurement does not depend on a model of Galactic diffuse emission, it has much smaller systematic errors than the DGRB intensity measurement.

Here, we consider the consistency of blazars with the DGRB intensity and anisotropy. 
We begin by describing our general blazar model, an LDDE+SED-sequence model similar to that in ABH. 
By constraining this model with the blazar $dN/dF$ and the DGRB anisotropy, we place limits on the LDDE model parameters. 
We place an upper limit on the contribution of blazars to the DGRB flux and predict the resolution of blazars by the Fermi-LAT at the end of 5 years of observations. 
We also place a lower limit on the blazar contribution to the DGRB anisotropy, which places a stringent constraint on all other sources of the DGRB intensity. 
Throughout the paper, we take a flat universe with the cosmological parameters $\Omega_{m}=0.272$, $\Omega_{\Lambda}=0.728$, and $H_0=70.2\rm\ km\ s^{-1}\ Mpc^{-1}$~\cite{Komatsu:2010fb}. 
{\it Note,} the use of $h$ in the text refers to Planck's constant, and not the Hubble parameter.

\begin{table}
\begin{center}
\begin{tabular}[t]{|l|l|l|}
  \multicolumn{3}{c}{Table~\ref{CPtable}} \\
  \hline
  $E\ ({\rm GeV})$ & Blazar $C_P$ & DGRB $C_P$\\
  \hline
  1.04-1.99 & $4.52\times10^{-18}$ & $4.62\times10^{-18}$\\
  1.99-5.00 & $1.29\times10^{-18}$ & $1.30\times10^{-18}$\\
  5.00-10.4 & $8.89\times10^{-20}$ & $8.45\times10^{-20}$\\
  10.4-50.0 & $2.36\times10^{-20}$ & $2.11\times10^{-20}$\\
  \hline
\end{tabular}
\caption{Comparison between the DGRB anisotropy $C_P$ for our best-fit blazar model, given the constraints from the observed blazar source-count distribution and DGRB anisotropy, and, for comparison, those observed in FA12. 
Values of $C_P$ are given in ${\rm (cm^{-2}s^{-1}sr^{-1})^2 sr}$.\label{CPtable}}
\end{center}
\end{table}

\section{The LDDE+SED-Sequence Model of Blazar Intensity and Anisotropy}

\subsection{The Blazar Gamma-ray Luminosity Function and SED-sequence}

The full model of blazars can be fully determined by two functions: the gamma-ray luminosity function (GLF) gives the comoving blazar density per unit luminosity and the SED gives the energy spectra of individual blazars. 
For a known redshift $z$ and blazar luminosity $L_{\gamma}$ (defined as $\nu L_{\nu}$ at $h\nu=100\rm\ MeV$), the GLF is $\rho_{\gamma}(L_{\gamma},z)$ and the SED is $\nu L_{\nu}(\nu,L_{\gamma})$. 
In this work, $\nu$ denotes the blazar rest-frame frequency.

By considering the blazar evolution as a function of both luminosity and redshift, we are effectively including both the FSRQ population, which has higher luminosity, and the BL Lac population, which has lower luminosity. 
Because FSRQs have higher luminosity, and therefore greater flux, than BL Lacs, the majority of unresolved blazars are BL Lacs. However, by considering all possible blazar luminosities and redshifts, we are including the contributions of both unresolved FSRQs and unresolved BL Lacs to the DGRB. 
The evolutionary properties of only the resolved FSRQs have been studied~\cite{Ajello:2011zi}, though these blazars have been resolved and therefore do not contribute to the DGRB. 

\begin{figure}[t]
\begin{center}
\includegraphics[width=5truein]{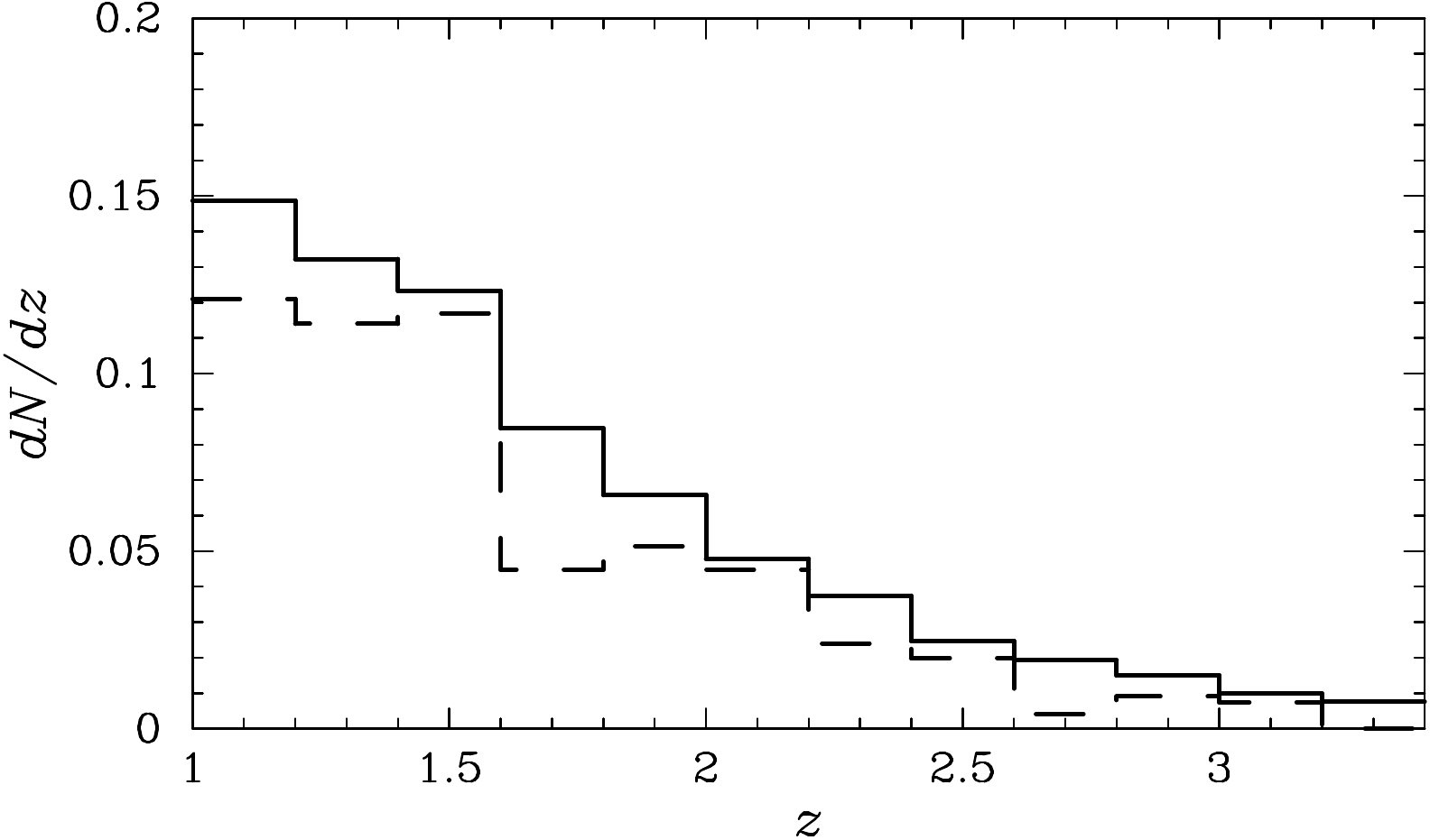}
\end{center}
\caption{Shown is the best fit model line for all 1FGL-observable blazars, both FSRQs and BL Lacs (solid line), relative to a measurement for the FSRQ population (dashed line)~\cite{Abdo:2010ge}. The consistency of the distributions is confirmed by a KS test, giving a probability of 78.6\% that
the distributions are consistent.
Each distribution has been normalized to unity.
\label{zdist}.}
\end{figure}

Above a redshift of unity, the contribution of BL-Lacs to the blazar redshift distribution is negligible, so the two populations are expected to be the same. For redshifts greater than unity, our model successfully reproduces the redshift
distribution of FSRQs in Ref.~\cite{Abdo:2010ge}, as is shown in figure~\ref{zdist}. A Kolmogorov-Smirnov test (KS test) gives a probability of 78.6\% that
the distributions are consistent, indicating that the model reproduces
the data well.

Because blazars and X-ray AGN are believed to be a related source population, observed at different projected angles, gamma-ray blazars should evolve similarly to X-ray AGN.
For blazars, the gamma-ray jet is being observed, whereas for X-ray AGN, the X-ray emission from the disk is being observed. 
Hard X-ray AGN follow an X-ray luminosity function (XLF) which has been parameterized by Ref.~\cite{Ueda:2003yx}. 
The XLF model of Ref.~\cite{Ueda:2003yx} is a luminosity-dependent density evolution model, which has a peak evolution redshift which depends on luminosity, so AGN of different luminosities will evolve differently.

The gamma-ray jet luminosity can be related to the X-ray disk luminosity $L_{X}$ as $P=10^{q}L_{X}$, where $q$ is a scaling parameter and $P=\int L_{\nu}d\nu$ is the bolometric blazar jet luminosity~\cite{Inoue:2008pk}. 
The bolometric luminosity from a blazar jet is proportional to the mass accretion rate $\dot{m}$. 
For blazars with high accretion rates close to the Eddington limit, AGN X-ray disk luminosity also goes as $L_{X}\propto\dot{m}$~\cite{Falcke:2003ia,Merloni:2003aq,Gallo:2005tf}. 
Because black hole growth primarily takes place near the Eddington limit~\cite{Marconi:2003tg}, the bolometric jet luminosity and the X-ray disk luminosity are proportional. If black hole growth were to take place far from the Eddington limit, the relative strengths of the bolometric jet luminosity and X-ray disk luminosity would vary~\cite{Ghisellini:2009rb}.

In addition to the luminosity scaling parameter $q$, the GLF may have a different faint-end luminosity index $\gamma_1$ than the XLF. 
Also, the small opening angle of the blazar jet means that most AGN are not observed as blazars, so the fraction of AGN observed as blazars is parameterized as $\kappa$. 
Note that this model does not require all AGNs to have blazar-like jets, but that the blazar population approximately follows that of the full AGN population, and is some fraction of it.  In turn, a fraction of those can be seen as blazars.
The comoving number density per unit $L_{\gamma}$ of gamma-ray blazars with luminosity $L_{\gamma}$ at redshift $z$ is then given as
\begin{equation}
\rho_{\gamma}(L_{\gamma},z)=\kappa\frac{dL_{X}}{dL_{\gamma}}\rho_{X}(L_{X},z)\enspace,
\label{rhogamma}
\end{equation}
where $\rho_{X}$ is the XLF. 
The precise relationship between X-ray AGN and gamma-rays blazars is not yet known. 
We are using the simple ansatz that they are related as shown in Eq.~\eqref{rhogamma}, as proposed by Inoue and Totani~\cite{Inoue:2008pk}, and using physical data on gamma-ray blazars to constrain the three parameters. 
To the best of our knowledge, this model satisfies all current observations and constraints, and is therefore a viable possibility.

The blazar SED sequence used in this work is that of Ref.~\cite{Fossati:1997vu,Fossati:1998zn,Donato:2001ge}. 
This SED bins blazars by radio luminosity, giving $\nu L_{\nu}$ over nineteen orders of magnitude in energy for each of five radio luminosity bins. 
The blazar SED shows two peaks in the blazar luminosity: a peak in the radio associated with synchrotron emission and a peak in the gamma-ray associated with inverse Compton emission. 
A shift in these peaks with blazar luminosity indicates that the gamma-ray spectrum becomes harder for lower luminosity blazars. 
This has been observed by noting that FSRQs have high luminosity and soft spectra, while BL Lacs have harder spectra with lower luminosities~\cite{Ghisellini:2009rb,Abdo:2009iq,Collaboration:2010yg,Lott:2012tp}. 
The importance of this spectral hardening at low luminosities cannot be overstated: the gamma-ray spectra of unresolved, lower-luminosity blazars should be harder than the spectra of resolved blazars. 

To parametrize the SED, we use the model of Ref.~\cite{Inoue:2008pk}, which fits the SED sequence of Ref.~\cite{Fossati:1997vu,Fossati:1998zn,Donato:2001ge}. 
With this SED model, we can relate a blazar's bolometric luminosity to its gamma-ray luminosity to estimate the average energy spectrum for a blazar of that luminosity. 
The fully parameterized model of the SED may be found in Appendix 1 of ABH.

\subsection{LDDE Model Calculation of Blazar Number, Intensity, and Anisotropy}

For a blazar at redshift $z$ with bolometric luminosity $P$, the observed gamma-ray flux in the energy band from $E_0^{\rm min}$ to $E_0^{\rm max}$ is
\begin{eqnarray}\label{Sflux}
  S(z,P)=&&\frac{1+z}{4\pi d_{L}(z)^{2}}\int_{(E_0^{\rm
      min})(1+z)/h}^{(E_0^{\rm
      max})(1+z)/h}d\nu\frac{L_{\nu}(\nu,P)}{h\nu}\ \exp\left[-\tau\left(z,\frac{h\nu}{1+z}\right)\right]\enspace,
\end{eqnarray}
where $d_{L}$ is the luminosity distance. 
The $\exp[-\tau(z,E_0)]$ factor accounts for photon absorption before reaching Earth. 
For our work, we choose the absorption factor of Gilmore et al.~\cite{Gilmore:2009zb}, which has determined lower values of opacity than previous estimates and is consistent with the findings of Ref.~\cite{Abdo:2010kz}. 
For the special case of the blazar flux above 100 MeV ($E_0^{\rm min}=100\rm\ MeV$ and $E_0^{\rm max}=\infty$) we refer to the flux as $F_{100}$. 

For a given GLF, the number count of blazars above a sensitivity $F_{\gamma}$ is
\begin{equation}\label{NgtF}
  N(>F_{\gamma})=4\pi\int_{0}^{z_{\rm max}}dz\frac{dV}{dz}\int_{L_{\gamma}^{\rm lim}(z,F_{\gamma})}^{\infty}dL_{\gamma}\rho_{\gamma}(L_{\gamma},z)\enspace,
\end{equation}
where $L_{\gamma}^{\rm lim}$ is the gamma-ray luminosity below which a blazar at redshift $z$ is no longer detectable for the sensitivity $F_{\gamma}$.
We set the parameter $z_{\rm max}=5$, but this does not affect the calculation significantly, since the GLF peak is at redshift of order unity.

Integrating the flux of an individual blazar, Eq.~\eqref{Sflux}, over all unresolved blazars, the total diffuse flux coming from unresolved blazars is
\begin{eqnarray}
  \frac{dN}{dE_{0}dAdtd\Omega}=&&\frac{1}{4\pi}\int_{0}^{z_{\rm
      max}}dz\frac{d\chi}{dz}e^{-\tau(z,E_0)}
  \int_{L_{\gamma,\rm min}}^{L_{\gamma}^{\rm lim}(F_{\gamma},z)}dL_{\gamma}\frac{\rho_{\gamma}(L_{\gamma},z)}{h}
  \frac{L_{\nu}[E/h,P(L_{\gamma})]}{E}\enspace.
\end{eqnarray}
Here, $E$ is the emitted photon energy, and $E_0 =
  E/(1+z)$ is the observed photon energy at Earth. $A$ is area
on Earth, $t$ is time on Earth, and $\Omega$ is solid angle in the sky. 
The integral over the comoving distance $\chi$ accounts for the line-of-sight of the observation. 
Because the luminosity integral diverges if $\gamma_1>1$, we place a lower bound of $L_{\gamma,\rm min}=10^{42}\rm\ erg\ s^{-1}$ on this integral. 
This is an order of magnitude lower luminosity than any Fermi-LAT observed blazar~\cite{Abdo:2009iq,Collaboration:2010yg}. 
That is, we impose a step-function cutoff of blazar GLF at $L_{\gamma,\rm min}$. 
However, our best-fit value has $\gamma_1<1$, so the final result is not strongly dependent on this cutoff. 

The intensity per solid angle of the DGRB in the energy band from $E_0^{\rm min}$ and $E_0^{\rm max}$ is given by
\begin{equation}
I=\int_{E_0^{\rm min}}^{E_0^{\rm max}}dE_0\frac{dN}{dE_{0}dAdtd\Omega}\enspace.
\end{equation}
The Poisson term of the blazar anisotropy per unit solid angle, $C_P$, is the square of the blazar flux, integrated over all unresolved blazars~\cite{Ando:2006cr,Cuoco:2012yf}. 
For the blazar SED+GLF model, this is given by
\begin{equation}
C_P=\int_{0}^{z_{\rm max}}dz\frac{d^2V}{dzd\Omega}\int_{L_{\gamma,\rm min}}^{L_{\gamma}^{\rm lim}(F_{\gamma},z)}dL_{\gamma}\rho_{\gamma}(L_{\gamma},z)S^2\enspace,
\end{equation}
with $V$ the comoving volume and $S$ the single blazar flux in a particular energy band, from Eq.~\eqref{Sflux}. 
This equation can be written as
\begin{eqnarray}
C_P=&&\int_{0}^{z_{\rm max}}dz\frac{d\chi}{dz}\int_{L_{\gamma,\rm min}}^{L_{\gamma}^{\rm lim}(F_{\gamma},z)}dL_{\gamma}\frac{\rho_{\gamma}}{d_L^2}\nonumber\\
&&\times\left[\int_{(E_0^{\rm min})(1+z)/h}^{(E_0^{\rm max})(1+z)/h} \frac{d\nu}{\nu}\frac{L_{\nu}(\nu,P)\exp\left[-\tau\left(z,h\nu/(1+z)\right)\right]}{4\pi h}\right]^2\enspace,
\end{eqnarray}
for luminosity distance $d_L$ and comoving distance $\chi$.

To calculate the threshold between observable and unobservable blazars, $L_{\gamma}^{\rm lim}$, we use the index-dependent flux limit shown in figure 1 of FB10. 
This plot gives the flux limit as a function of blazar index for test statistic $TS=25$, which is consistent with the DGRB flux~\cite{Abdo:2010nz} and anisotropy~\cite{Ackermann:2012uf} analyses, which were done using the full Fermi-LAT 1-year catalog (1FGL). 
The blazar $dN/dF$ of FB10 is tabulated for $TS=50$ in the 1FGL, so for that calculation we use the flux limit shown in figure 1 of ABH.

Recently, Ref.~\cite{Singal:2011yi} has suggested another approach to determining the true source population of blazars, to account for the observational biases discussed above. 
In that analysis, the correlation between the observed blazar flux and the observed blazar index distribution was estimated. 
By removing this observational bias, they determined the true blazar index distribution and $dN/dF$ for the observed Fermi-LAT blazars. 
This method uses the index and flux of a given blazar to determine whether it is ``resolved'' or ``unresolved,'' rather than the test statistic $TS$ that has an intrinsically higher value for harder Fermi-LAT sources.
In future analyses, such a technique may be used to limit the observational bias of the observed blazar source population. Using this method, Ref.~\cite{Singal:2011yi} calculates an intensity of $I(>100{\rm\ MeV})=7.7^{+0.8}_{-1.2}\times10^{-6}{\rm\ ph\ cm^{-2}\ s^{-1}\ sr^{-1}}$ for blazars, which is significantly larger than the value determined in this work when considering the DGRB anisotropy limits.
In this work, we calculate the Fermi-LAT sensitivity threshold based on the $TS$ values of FB10, as discussed above.

\section{Results for Inferred Relation Between Blazar Intensity and Anisotropy}

Our adopted LDDE gamma-ray blazar model based on the X-ray AGN XLF requires three free parameters: $q$, $\gamma_1$, and $\kappa$. 
All other parameters in the blazar model are fixed by other data, such as the AGN XLF and blazar SED sequence. 
\begin{figure}[t]
\begin{center}
\includegraphics[width=5truein]{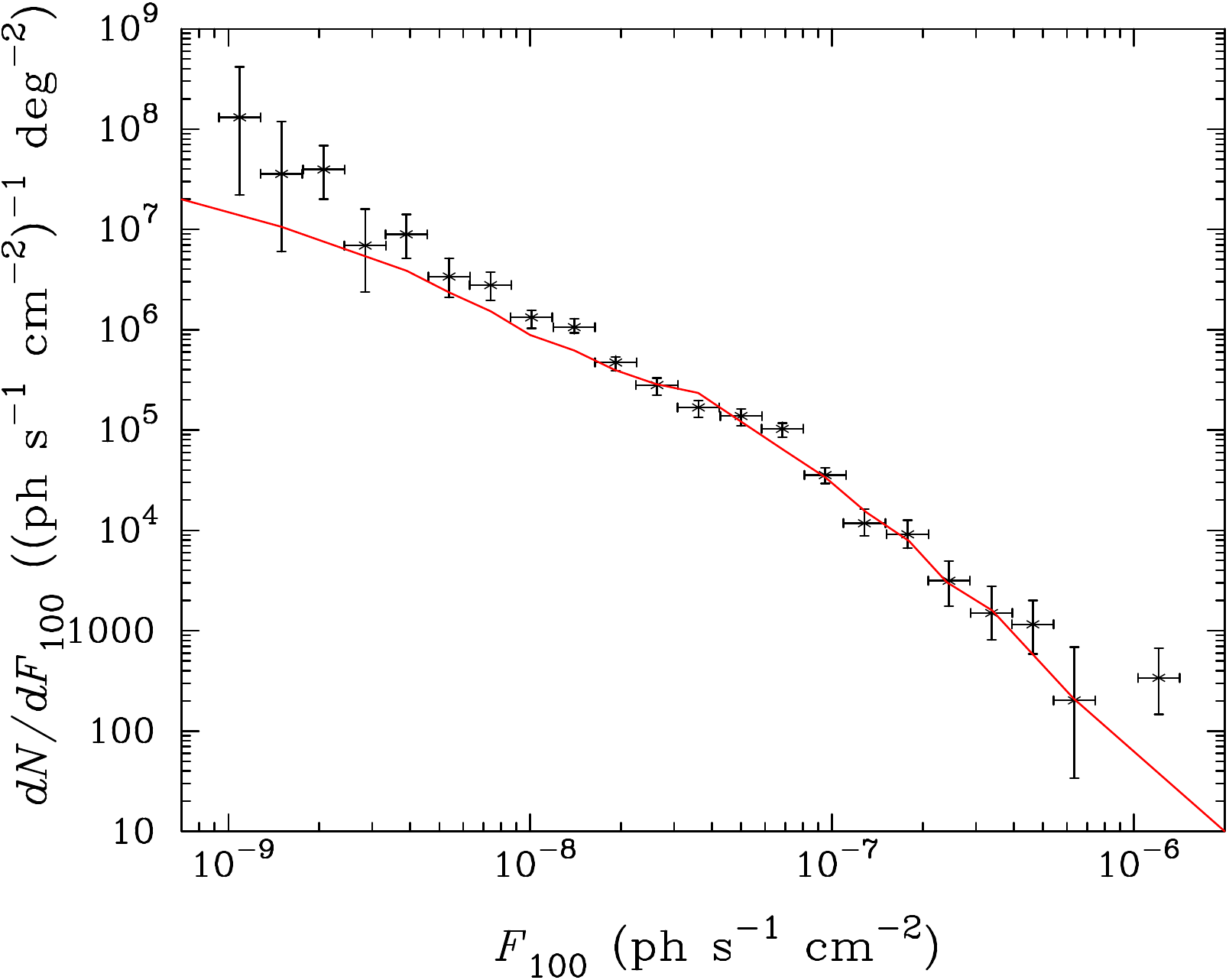}
\end{center}
\caption{Shown is the best-fit model for the source-count distribution function $dN/dF$ (solid line), given the constraints from the observed blazar source-count distribution and DGRB anisotropy.  
The data are from FB10.\label{fitteddNdF}}
\end{figure}

First, we consider the consistency between the blazar $dN/dF$ of FB10 with both the DGRB spectrum from Ref.~\cite{Abdo:2010nz} and the DGRB anisotropy from FA12. 
For no values of $q$, $\gamma_1$, and $\kappa$ could the blazars account for both the full DGRB flux and DGRB anisotropy. 
In particular, values of the model which were consistent with the DGRB flux at high energies overproduced the DGRB anisotropy by a factor of $\sim5$. 
The DGRB anisotropy constrains the blazar population much more than the DGRB intensity does. 
We also considered a model which was fit only to the blazar $dN/dF$. 
The best-fit model for this fit also overproduced the DGRB anisotropy, by a factor of 3.6. 
Therefore, the DGRB anisotropy itself places a significant constraint on the LDDE+SED-sequence blazar model.

Considering only the consistency between the blazar $dN/dF$ of FB10 with the DGRB anisotropy from FA12, we found that the DGRB anisotropy can be attributed entirely to unresolved blazars. 
The best-fit parameter values and errors to match the anisotropy and $dN/dF$ are $q=4.21^{+0.16}_{-0.09}$, $\gamma_{1}=0.43^{+0.14}_{-0.07}$, and $\log_{10}(\kappa/10^{-6})=0.98^{+0.14}_{-0.12}$. 
The model reproduces well the blazar $dN/dF$ and the DGRB $C_P$, with a reduced $\chi^2/{\rm DOF} = 0.76$ for the combined data. To calculate the $\chi^2$, we have assumed that each data point in the DGRB anisotropy and each data point in the blazar $dN/dF$ is independent, so we have included all 26 data points from the two data sets in our calculation.
With an adopted Fermi-LAT 5-year sensitivity of $F_{100}=2\times10^{-9}\rm\ ph\ cm^{-2}\ s^{-1}$~\cite{Atwood:2009ez}, we expect $1631^{+87}_{-345}$ blazars to be resolved, comprising $98.7\pm 0.1$ percent of the total (resolved plus unresolved) blazar flux. The parameter values given by matching the $dN/dF$ and anisotropy are significantly different from previous work which had found consistency between blazars and the DGRB flux (ABH and Ref.~\cite{Inoue:2008pk}), yet those models overproduce the DGRB anisotropy by a factor of $\sim5$ (e.g. $q=4.19$, $\gamma_{1}=1.51$, and $\log_{10}(\kappa/10^{-6})=0.38$ from ABH). 

The value of $q$ indicates that the bolometric luminosity of a blazar jet is roughly $\sim$16,000 times more luminous than the X-ray from the accretion disk. 
The low value for $\gamma_1$ indicates that the anisotropy and flux due to blazars is dominated by high-luminosity blazars, as would be expected. 
The blazar fraction $\kappa$ indicates that of every $\sim$105,000 AGN, one is observed as a blazar. 
If one assumes that every AGN is a blazar, this corresponds to an opening angle of $\sim 0.6\rm\ deg$ for the average blazar jet, similar to the intrinsic jet opening angle of $\sim$ 1 deg~\cite{Pushkarev:2009dx}. 
Our fit to the DGRB anisotropy is given in Table~\ref{CPtable}.

The $dN/dF$ for our best-fit parameters is shown in figure~\ref{fitteddNdF}. 
At low fluxes, the $dN/dF$ flattens out instead of continuing upward to zero flux. 
This differs significantly from the broken power-law $dN/dF$ assumed by much previous work.

Given the blazar LDDE+SED-sequence model which best fits the blazar $dN/dF$ and the DGRB anisotropy, we can infer the resulting constraints these measurements place on the flux of unresolved blazars. 
For the best-fit model, blazars only contribute $5.9^{+2.1}_{-1.0}\%$ (68\% CL) of the DGRB intensity above 1 GeV. 
Moreover, the blazar flux for our best-fit model does not produce greater than 8\% of the DGRB flux in any single energy bin (68\% CL). 
Above 100 MeV, the best-fit model gives an intensity above 100 MeV of that is
$4.0^{+1.9}_{-1.2}\%$ (68\% CL) of the DGRB. This is less than the value from
Ref.~\cite{Ajello:2011zi}, which predicts the intensity above 100 MeV to be 9.3\% of
the DGRB from FSRQs alone.

\begin{figure}[t]
\begin{center}
\includegraphics[width=5truein]{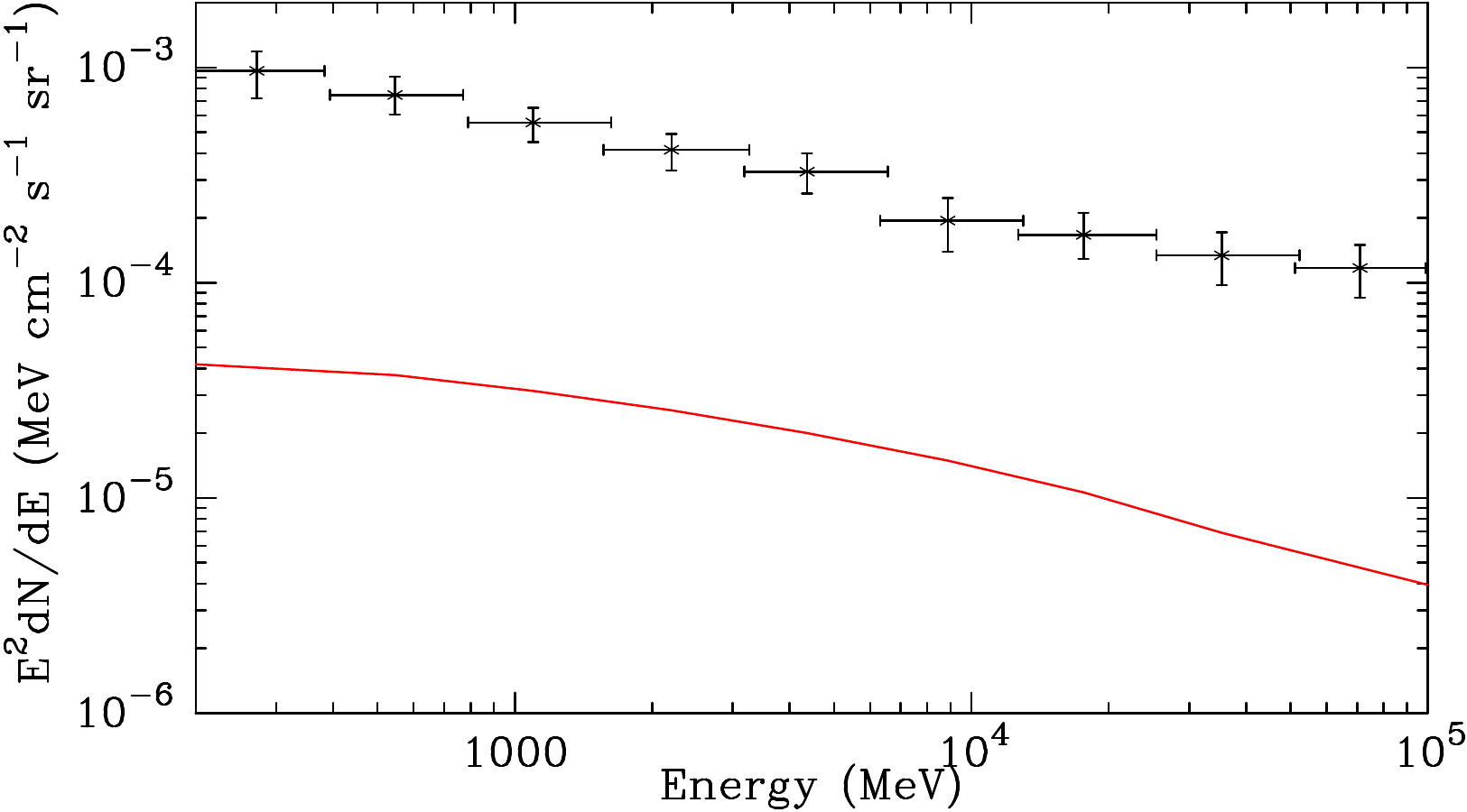}
\end{center}
\caption{The best-fit model for the total blazar intensity (solid red line), given the constraints from the observed blazar source-count distribution and DGRB anisotropy. 
The DGRB data are from Ref.~\cite{Abdo:2010nz}.\label{DGRBfitandpredict}}
\end{figure}

The spectrum of the inferred contribution of blazars to the DGRB relative to that observed is shown in figure~\ref{DGRBfitandpredict}.

As discussed above, the best-fit blazar model is consistent with the full DGRB anisotropy. 
While it is clear that a valid blazar model cannot overproduce the DGRB anisotropy, it is possible that blazars are not a primary source of this anisotropy. 
To test this, we ran a further test of how little of the anisotropy is due to blazars. 
In this test, we included the three free parameters discussed above as well as a fourth free parameter which accounted for the fraction of the total measured $C_P$ that is due to blazars. 
We found that in this test, the best-fit values of $q$, $\gamma_1$, and $\kappa$ were similar to the values discussed above for the combined $dN/dF$ and DGRB anisotropy fit. 
However, we did find that the best-fit value of the blazar $C_P$ was approximately the total DGRB $C_P$ (98\% of the total). 
This value of blazar $C_P$ is consistent with the full DGRB anisotropy (68\% CL), as is our best-fit model discussed above. 
In addition, this test finds that greater than 76\% of the full DGRB $C_P$ is due to blazars (68\% CL). 
This provides a strong limit on any other potential contributor to the DGRB flux: any additional population which contributes to the remaining 92\% of the DGRB flux must produce less than 24\% of the DGRB anisotropy.

The anisotropy-constrained LDDE+SED-sequence blazar model contributes much less of the intensity of the DGRB than in ABH. ABH had previously found consistency between the blazar $dN/dF$ and the full DGRB intensity; here, blazars are constrained to less than 5.9\% of the DGRB intensity above 1 GeV. 
Though the model parameters  $q$, $\gamma_1$, and $\kappa$ in ABH allowed for a blazar model which was consistent with the full DGRB intensity, that model also severely overproduced the DGRB anisotropy. 
The DGRB anisotropy constraint changed the best-fit parameters significantly --- e.g., the blazar faint-end index $\gamma_1$ changed from 1.51 (ABH) to 0.43 when the DGRB anisotropy was considered. 
This illustrates the importance of the DGRB anisotropy is constraining populations' contributions to the DGRB intensity.

\section{Discussion and Comparison to Previous Results}
\subsection{Power-Law Extrapolation Models}

The blazar source-count distribution has been historically modeled by as a broken power-law in $dN/dF$ when little was known regarding AGN population evolution and the background cosmological model.
This simplifying assumption was adopted  in FB10 and CKS. 
In such work, the broken power-law is fit to the source count distribution function above the point-source sensitivity observed by the Fermi-LAT. 
Below the point source resolution threshold, the faint-end power-law of this $dN/dF$ continues to rise to zero flux, which produces the unphysical result of an infinite number of blazars within our observable horizon.
To avoid this divergence, the $dN/dF$ is expected to flatten at low flux, which occurs naturally in physically-derived models such as the LDDE+SED-sequence model.

The spectra of individual blazars were approximated in FB10 as a distribution of single power-laws in energy.                                        
It was, however, shown in Refs.~\cite{Abdo:2009iq,Collaboration:2010yg} that a single power-law spectrum does a very poor job of fitting individual blazars, which display curvature and breaks in their spectra that are correlated with their total luminosity.
Ref.~\cite{Fossati:1997vu,Fossati:1998zn,Donato:2001ge} show that blazar gamma-ray spectra are soft for high-luminosity blazars and get much harder for low-luminosity blazars, and the wide variation of blazar spectra has been confirmed by Fermi-LAT~\cite{Collaboration:2010yg}.
The correlation of luminosity with spectrum leads to a hardening of the blazar spectra below threshold with respect to the spectrum above threshold. 
By not including a flux-dependence in the blazar index distribution, FB10 has assumed a softer blazar spectrum below threshold than is expected from observations of low-luminosity blazars.

Rather than considering the spectra of individual blazars, CKS considers only the aggregate blazar flux with a fixed power law in energy, which is inconsistent with the observed spectral variability of blazars and their dependence on luminosity~\cite{Fossati:1997vu,Fossati:1998zn,Donato:2001ge,Collaboration:2010yg,Meyer:2011uk}.
This simplification can drastically affect the results, since the Fermi-LAT threshold flux significantly varies for blazars with spectral index, e.g, as in figure 1 of FB10. 
Therefore, a common flux threshold, as assumed in CKS, for all blazars independent of spectral index is an unnecessary and inaccurate approximation. 
The blazar intensity is most sensitive to unresolved blazars with hard spectra, including those with fluxes somewhat below threshold;
conversely, the anisotropy is sensitive only to the blazar flux and is therefore most sensitive to blazars with flux just below and directly at the threshold. 
For an index-independent calculation of the flux threshold, the threshold used to calculate the intensity should be at a lower flux than that for the anisotropy calculation.
Simple power-law blazar models such as those in CKS and FB10 have been historically used to extrapolate the blazar population below to fluxes below current observations. 
However, due to the reasons given above and elaborated on in Ref.~\cite{Harding:2012sa}, these extrapolation models can introduce significant unjustified approximations below the observational Fermi-LAT threshold and the behavior of the population near the threshold.

\subsection{Differences in Results Between the LDDE+SED Model and Power-Law Models}

\begin{figure}[t]
\begin{center}
\includegraphics[width=5truein]{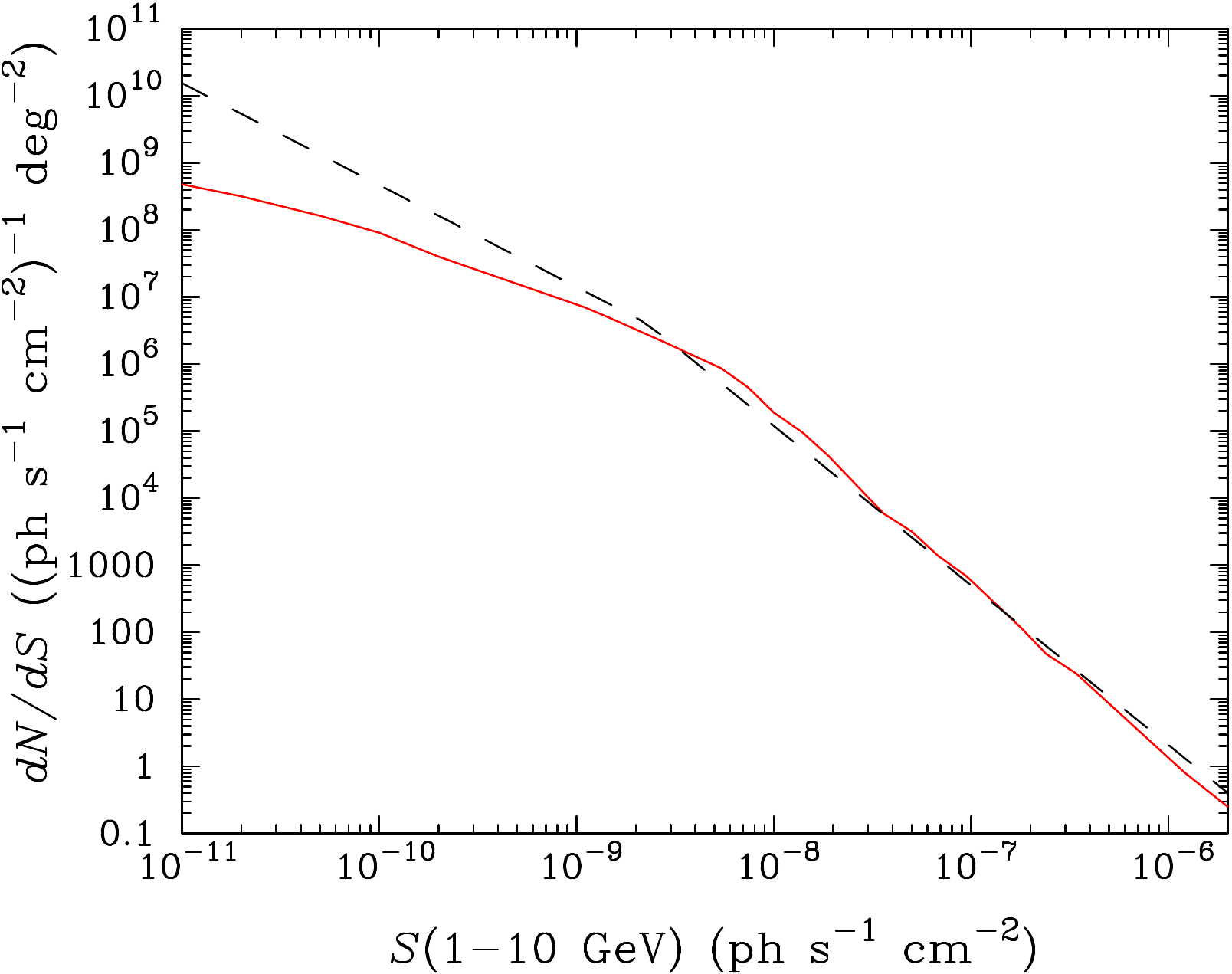}
\end{center}
\caption{Shown is the best-fit model for the source-count distribution function $dN/dS$ for flux $S$ as in Eq.~\eqref{Sflux} with $E_0^{\rm min}=1\rm\ GeV$ and $E_0^{\rm max}=10\rm\ GeV$ (solid red line), given the constraints from the observed blazar source-count distribution and DGRB anisotropy. 
For comparison, the best-fit $dN/dS$ from FB10, used in CKS, is also shown (dashed black line). Note that at low fluxes, the two extrapolated models differ significantly.\label{dNdS110GeV}}
\end{figure}

The differences between the calculations using the LDDE+SED-sequence blazar model considered here and the power-law models are threefold: the LDDE+SED-sequence gives a $dN/dF$ which flattens below the Fermi-LAT threshold, the LDDE+SED-sequence model accounts for the hardening of the average blazar spectra for low-flux blazars, and the LDDE+SED-sequence model calculates an index-dependent Fermi-LAT threshold flux based on the harder low-flux blazars. 
Below $F_{100}\approx 4\times10^{-9}\rm\ ph\ cm^{-2}\ s^{-1}$, the blazar $dN/dF$ for the LDDE+SED-sequence model rises less steeply than the broken power-law model of FB10. 
For comparison to CKS, we calculate $dN/dS$ with the flux from 1-10 GeV, $S$, calculated as in Eq.~\eqref{Sflux} with $E_0^{\rm min}=1\rm\ GeV$ and $E_0^{\rm max}=10\rm\ GeV$. 
The comparison between these two models can be seen in figure~\ref{dNdS110GeV}. 
Although at high fluxes, where the Fermi-LAT data is well-constrained, the two models give similar source-count distributions, the LDDE+SED-sequence model flattens out at low flux instead of continuing to rise down to zero flux. 
This is expected in physically motivated models, which have a continuous roll-off with flux rather than a hard break between two unrelated power-laws. 

\begin{figure}[t]
\begin{center}
\includegraphics[width=5truein]{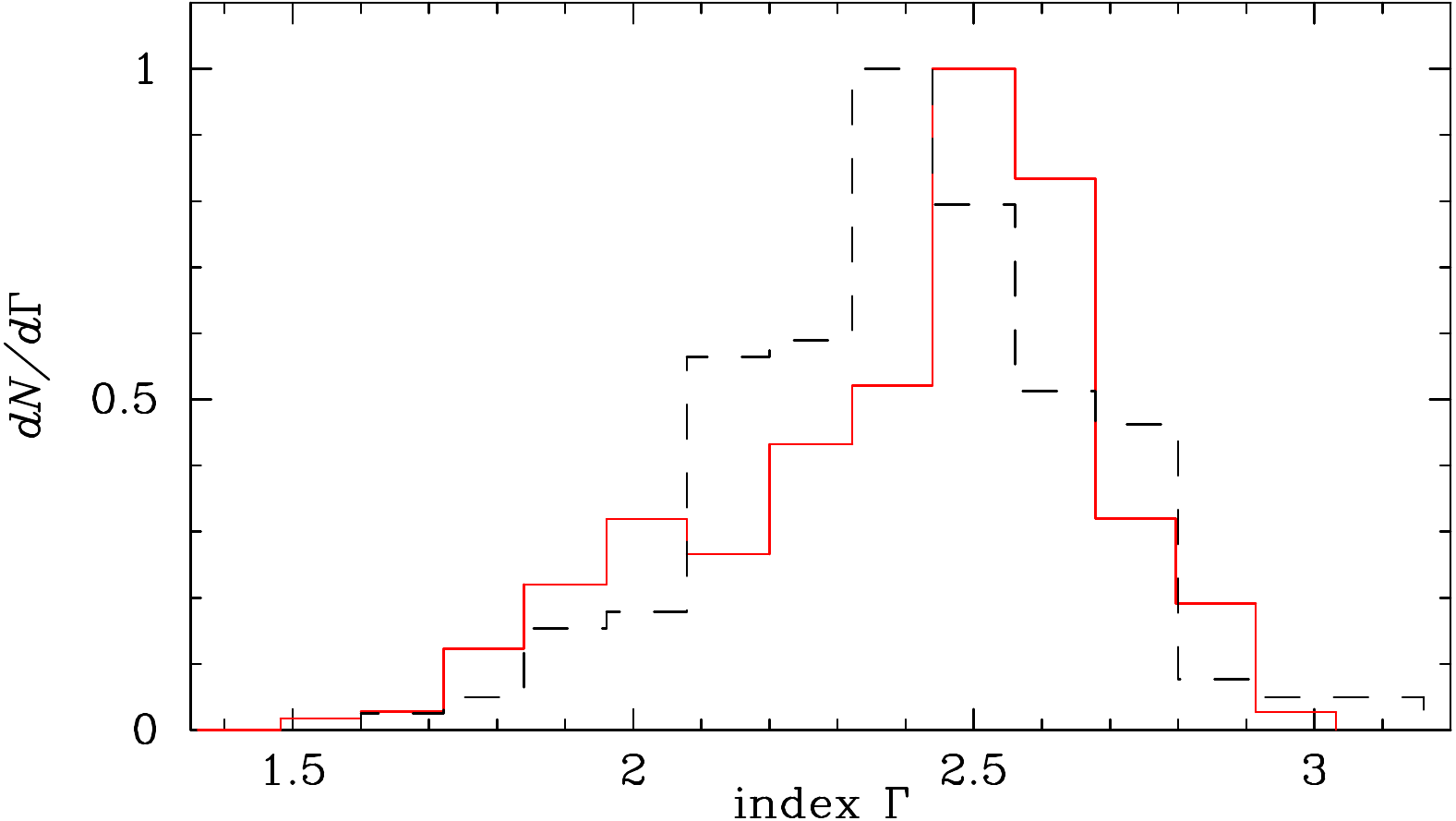}
\end{center}
\caption{Shown is the index distribution  in our best-fit model above $F_{100}=7\times10^{-8}\rm\ ph\ cm^{-2}\ s^{-1}$ (solid red line), given the constraints from the observed blazar source-count distribution and DGRB anisotropy.  
The measured index distribution above $F_{100}=7\times10^{-8}\rm\ ph\ cm^{-2}\ s^{-1}$ from FB10 is shown for comparison (dashed black line). 
A KS test gives a probability of 82.9\% that
the two distributions are consistent.
Both distributions in the figure have been normalized to unity.\label{indexdist}}
\end{figure}

The most striking feature of the LDDE+SED-sequence blazar model is due to the nature of the SED sequence itself. 
As blazar luminosity decreases, the peak of the inverse Compton emission shifts to higher energies. 
In the Fermi-LAT energy range, from 100 MeV to 100 GeV, this trend is observable as a hardening in the intrinsic blazar spectrum at low fluxes. 

Where the Fermi-LAT point-source resolution is nearly complete, above $F_{100}=7\times10^{-8}\rm\ ph\ cm^{-2}\ s^{-1}$, the LDDE+SED-sequence model matches the index distribution observed in FB10. This is shown in figure~\ref{indexdist} by comparing the blazar index distribution of the LDDE+SED-sequence model above $F_{100}=7\times10^{-8}\rm\ ph\ cm^{-2}\ s^{-1}$ to the measured blazar index distribution above the same threshold. A KS test gives a probability of 82.9\% that the two index distributions above $F_{100}=7\times10^{-8}\rm\ ph\ cm^{-2}\ s^{-1}$ are consistent.
However, FB10 assumes that this index distribution is identical at all sensitivities while the LDDE+SED-sequence model accounts for the hardening in the blazar spectra at low flux. 

Additionally, there is a dependence of the threshold flux on the blazar index due to the bias of the Fermi-LAT towards observations of hard sources. 
For fluxes above 100 MeV, $F_{100}$, the flux threshold varies by two orders of magnitude with blazar index. 
For higher energy fluxes, this bias is much weaker, but still varies somewhat. 
For the flux from 1-10 GeV, for example, we calculate a threshold range from $2.4-7.4\times10^{-10}{\rm ph\ cm^{-2}\ s^{-1}}$ for blazar indices from 1 to 1.5. 
These values are comparable to the value $3.6\times10^{-10}{\rm ph\ cm^{-2}\ s^{-1}}$ from FB10 or the value $3.7\times10^{-10}{\rm ph\ cm^{-2}\ s^{-1}}$ in CKS.

For a blazar model with a single power-law spectrum, the blazar spectrum has a similar slope to the DGRB flux and therefore the intensity is a measure of the blazar flux in any bin. 
However, unresolved blazars have lower flux and therefore are expected to have a harder spectrum than the resolved blazars. 
For a source population with a harder spectrum than the DGRB, intensity is largely a measure only of the flux in the lowest energy bin, not a true indication of the source's contribution to the DGRB. 
Consider, for instance, the blazar model of ABH. 
The blazar population of ABH contributes only $\sim20\%$ of the DGRB intensity because it has a much lower flux than the DGRB below 1 GeV. 
However, above 10 GeV, the ABH blazar model produces $100\%$ of the DGRB flux. 
Therefore, DGRB flux, binned in energy, is a better determination of whether a particular model fits the DGRB than the integrated intensity is.

Given the blazar LDDE+SED-sequence model which best fits the blazar $dN/dF$ and the DGRB anisotropy, the LDDE+SED-sequence model constrains the flux from blazars to only contribute $5.9^{+2.1}_{-1.0}\%$ (68\% CL) of the DGRB intensity above 1 GeV. 
This is a much stronger limit on the flux due to blazars than that from CKS or FA12, which derived limits of 30\% and 19\%, respectively. 
The blazar flux for the LDDE+SED-sequence best-fit model does not produce greater than 8\% of the DGRB flux in any single energy bin (68\% CL). 

\section{Conclusions}

We have shown that the DGRB anisotropy can be composed entirely by blazar emission when adopting an LDDE+SED-sequence model.  
Given the constraints from the observed DGRB anisotropy, the blazars are likely  only  a sub-dominant contribution to the DGRB flux.  
That is, blazars cannot simultaneously account for the DGRB intensity as well as anisotropy in the LDDE+SED-sequence model we study.
The LDDE+SED-sequence is an observationally-motivated physical-evolution model of the gamma-ray blazar population  which successfully accounts for the shape of the blazar $dN/dF$ below the current threshold, while taking into account the change in blazar spectra with flux, and the spectral dependence of the flux threshold. 
Models with a fixed source-count distribution power-law below a fixed Fermi-LAT point-source threshold fail to incorporate these crucial observed properties of the population.

Using the LDDE+SED-sequence model, we find that unresolved blazars can account for all of the DGRB anisotropy, in general agreement with the conclusions of CKS and FA12. 
In the LDDE+SED-sequence model, blazars are expected to be the dominant source of anisotropy in the DGRB, comprising greater than 76\% of the anisotropy (68\% CL). 
However, the constrained LDDE+SED-sequence blazar model produces only $5.7^{+2.1}_{-1.0}\%$ (68\% CL) of the total DGRB intensity and less than 8\% of the DGRB flux in any energy bin (68\% CL). 
This is significantly less intensity than that calculated using the power-law models of FB10 or CKS. 

Other populations which contribute to the DGRB are highly constrained by these results. 
Emission sources other than blazars, such as non-blazar AGN~\cite{Inoue08}, millisecond pulsars~\cite{FaucherGiguere09}, star-forming galaxies~\cite{Fields10,Stecker:2010di,Chakraborty:2012sh,Lacki:2012si,Ackermann:2012eba}, radio galaxies~\cite{Inoue:2011bm}, circum-galactic gas~\cite{Feldmann:2012rx}, or dark matter annihilation~\cite{Ando:2005xg,Ando:2006cr,Miniati07,Baltz08,SiegalGaskins:2008ge,SiegalGaskins:2009ux,Hensley09,Fornasa:2009qh,Abazajian:2010zb}, may be responsible for a significant portion of the DGRB flux, but such sources are constrained to provide less than 24\% of the DGRB anisotropy. 
For many sources, this anisotropy constraint limits the amount of DGRB intensity those sources can provide.

Recent studies have also shown that electromagnetic cascades from very high-energy emission of diffuse emission sources can contribute to the DGRB~\cite{Kneiske:2007jq,Venters:2010bq,Venters:2012bx,Inoue:2012cs}. In some cases, this electromagnetic cascade can be used to constrain a particular source class's contribution to the DGRB intensity and anisotropy. In future analyses, such a technique may be used to further limit the blazar contribution to the DGRB in the LDDE+SED-sequence model.

There are known deficiencies in the blazar LDDE+SED-sequence model's agreement with all currently observed blazar SEDs, and this must be addressed with further studies of the Fermi-LAT data. 
The blazar SED sequence considered here bases its gamma-ray spectral information on the few blazars observed by EGRET~\cite{Fossati:1997vu,Fossati:1998zn,Donato:2001ge}.  
A new SED-dependence on blazar luminosity relation is strongly motivated, based on the latest Fermi-LAT results~\cite{Abdo:2009iq,Collaboration:2010yg,Meyer:2011uk}, and would likely greatly increase the accuracy of the predictions physical blazar population models.  
However, the blazar LDDE+SED-sequence is the best developed blazar-evolution model that goes far beyond the overly simple assumptions of a single power law for the full blazar population with a broken power-law source-count distribution.

In summary, the LDDE+SED-sequence model of blazars is consistent with both the blazar source count distribution and the DGRB anisotropy. 
These results find that the blazar anisotropy limits blazars to be a sub-dominant contribution to the total DGRB flux.
Physically-motivated models of the blazar SED-sequence and the blazar GLF will inevitably lead to a better understanding of blazars, non-blazar AGN, and the nature the contributors of the diffuse extragalactic gamma-ray sky.

\begin{acknowledgments}
 JPH \& KNA are partially supported by NSF CAREER Grant No.\ 11-59224.
\end{acknowledgments}

\bibliography{bibliography}

\providecommand{\href}[2]{#2}\begingroup\raggedright\begin{thebibliography}{10}

\bibitem{collaboration:2011bg}
{\bf Fermi-LAT Collaboration} Collaboration, M.~Ackermann {\em et.~al.}, {\it
  {The Second Catalog of Active Galactic Nuclei Detected by the Fermi Large
  Area Telescope}},  \href{http://xxx.lanl.gov/abs/1108.1420}{{\tt
  arXiv:1108.1420}}.

\bibitem{Fossati:1997vu}
G.~Fossati {\em et.~al.}, {\it {Unifying models for X-ray selected and Radio
  selected BL Lac Objects}},  {\em \aap} {\bf 375} (Sept., 1997) 739--751,
  [\href{http://xxx.lanl.gov/abs/astro-ph/9704113}{{\tt astro-ph/9704113}}].

\bibitem{Fossati:1998zn}
G.~Fossati {\em et.~al.}, {\it {A Unifying View of the Spectral Energy
  Distributions of Blazars}},  {\em Mon. Not. Roy. Astron. Soc.} {\bf 299}
  (1998) 433--448, [\href{http://xxx.lanl.gov/abs/astro-ph/9804103}{{\tt
  astro-ph/9804103}}].

\bibitem{Donato:2001ge}
D.~Donato {\em et.~al.}, {\it {Hard X-ray properties of blazars}},  {\em \aap}
  {\bf 375} (Sept., 2001) 739--751,
  [\href{http://xxx.lanl.gov/abs/astro-ph/0105203}{{\tt astro-ph/0105203}}].

\bibitem{Blandford:1979za}
R.~D. {Blandford} and A.~{Konigl}, {\it {Relativistic jets as compact radio
  sources}},  {\em \apj} {\bf 232} (Aug., 1979) 34--48.

\bibitem{Dermer95}
C.~D. {Dermer} and N.~{Gehrels}, {\it {Two Classes of Gamma-Ray--emitting
  Active Galactic Nuclei}},  {\em \apj} {\bf 447} (July, 1995) 103--+.

\bibitem{Padovani93}
P.~{Padovani} {\em et.~al.}, {\it {Radio-loud AGN and the extragalactic
  gamma-ray background}},  {\em \mnras} {\bf 260} (Feb., 1993) L21--L24.

\bibitem{Stecker93}
F.~W. {Stecker}, M.~H. {Salamon}, and M.~A. {Malkan}, {\it {The high-energy
  diffuse cosmic gamma-ray background radiation from blazars}},  {\em \apjl}
  {\bf 410} (June, 1993) L71--L74.

\bibitem{Salamon94}
M.~H. {Salamon} and F.~W. {Stecker}, {\it {The blazar gamma-ray luminosity
  function and the diffuse extragalactic gamma-ray background}},  {\em \apjl}
  {\bf 430} (July, 1994) L21--L24.

\bibitem{Chiang95}
J.~{Chiang} {\em et.~al.}, {\it {The Evolution of Gamma-Ray--loud Active
  Galactic Nuclei}},  {\em \apj} {\bf 452} (Oct., 1995) 156--+.

\bibitem{Stecker96}
F.~W. {Stecker} and M.~H. {Salamon}, {\it {The Gamma-Ray Background from
  Blazars: A New Look}},  {\em \apj} {\bf 464} (June, 1996) 600--+,
  [\href{http://xxx.lanl.gov/abs/9601120}{{\tt 9601120}}].

\bibitem{Chiang98}
J.~{Chiang} and R.~{Mukherjee}, {\it {The Luminosity Function of the EGRET
  Gamma-Ray Blazars}},  {\em \apj} {\bf 496} (Mar., 1998) 752--+.

\bibitem{Mucke00}
A.~{M{\"u}cke} and M.~{Pohl}, {\it {The contribution of unresolved radio-loud
  AGN to the extragalactic diffuse gamma-ray background}},  {\em \mnras} {\bf
  312} (Feb., 2000) 177--193.

\bibitem{Giommi05}
P.~Giommi {\em et.~al.}, {\it {Non-thermal Cosmic Backgrounds from Blazars: the
  contribution to the CMB, X-ray and gamma-ray Backgrounds}},  {\em Astron.
  Astrophys.} {\bf 445} (2006) 843--855,
  [\href{http://xxx.lanl.gov/abs/astro-ph/0508034}{{\tt astro-ph/0508034}}].

\bibitem{Narumoto06}
T.~Narumoto and T.~Totani, {\it {Gamma-Ray Luminosity Function of Blazars and
  the Cosmic Gamma-Ray Background: Evidence for the Luminosity Dependent
  Density Evolution}},  {\em Astrophys. J.} {\bf 643} (2006) 81,
  [\href{http://xxx.lanl.gov/abs/astro-ph/0602178}{{\tt astro-ph/0602178}}].

\bibitem{Dermer07}
C.~D. {Dermer}, {\it {Statistics of Cosmological Black Hole Jet Sources: Blazar
  Predictions for the Gamma-Ray Large Area Space Telescope}},  {\em \apj} {\bf
  659} (Apr., 2007) 958--975, [\href{http://xxx.lanl.gov/abs/0605402}{{\tt
  0605402}}].

\bibitem{Pavlidou08}
V.~{Pavlidou} and T.~M. {Venters}, {\it {The Spectral Shape of the Gamma-Ray
  Background from Blazars}},  {\em \apj} {\bf 673} (Jan., 2008) 114--118,
  [\href{http://xxx.lanl.gov/abs/0710.0002}{{\tt arXiv:0710.0002}}].

\bibitem{Bhattacharya09}
D.~Bhattacharya, P.~Sreekumar, and R.~Mukherjee, {\it {Gamma-ray luminosity
  function of Gamma-ray bright AGNs}},  {\em Res. Astron. Astrophys.} {\bf 9}
  (2009) 85--94, [\href{http://xxx.lanl.gov/abs/0811.4388}{{\tt
  arXiv:0811.4388}}].

\bibitem{Collaboration:2010gqa}
{\bf Fermi-LAT Collaboration} Collaboration, A.~A. Abdo {\em et.~al.}, {\it
  {The Fermi-LAT high-latitude Survey: Source Count Distributions and the
  Origin of the Extragalactic Diffuse Background}},  {\em Astrophys. J.} {\bf
  720} (2010) 435--453, [\href{http://xxx.lanl.gov/abs/1003.0895}{{\tt
  arXiv:1003.0895}}].

\bibitem{Abazajian:2010pc}
K.~N. Abazajian, S.~Blanchet, and J.~Harding, {\it {The Contribution of Blazars
  to the Extragalactic Diffuse Gamma-ray Background and Their Future Spatial
  Resolution}},  {\em Phys.Rev.} {\bf D84} (2011) 103007,
  [\href{http://xxx.lanl.gov/abs/1012.1247}{{\tt arXiv:1012.1247}}]. (ABH).

\bibitem{Stecker:2010di}
F.~W. Stecker and T.~M. Venters, {\it {Components of the Extragalactic Gamma
  Ray Background}},  {\em Astrophys.J.} {\bf 736} (2011) 40,
  [\href{http://xxx.lanl.gov/abs/1012.3678}{{\tt arXiv:1012.3678}}].

\bibitem{Malyshev:2011zi}
D.~Malyshev and D.~W. Hogg, {\it {Statistics of gamma-ray point sources below
  the Fermi detection limit}},  {\em Astrophys.J.} {\bf 738} (2011) 181,
  [\href{http://xxx.lanl.gov/abs/1104.0010}{{\tt arXiv:1104.0010}}].

\bibitem{Ueda:2003yx}
Y.~Ueda, M.~Akiyama, K.~Ohta, and T.~Miyaji, {\it {Cosmological Evolution of
  the Hard X-ray AGN Luminosity Function and the Origin of the Hard X-ray
  Background}},  {\em Astrophys. J.} {\bf 598} (2003) 886--908,
  [\href{http://xxx.lanl.gov/abs/astro-ph/0308140}{{\tt astro-ph/0308140}}].

\bibitem{Hasinger05}
G.~Hasinger, T.~Miyaji, and M.~Schmidt, {\it {Luminosity-dependent evolution of
  soft X-ray selected AGN: New Chandra and XMM-Newton surveys}},  {\em Astron.
  Astrophys.} {\bf 441} (2005) 417--434,
  [\href{http://xxx.lanl.gov/abs/astro-ph/0506118}{{\tt astro-ph/0506118}}].

\bibitem{Cuoco:2012yf}
A.~Cuoco, E.~Komatsu, and J.~Siegal-Gaskins, {\it {Joint anisotropy and source
  count constraints on the contribution of blazars to the diffuse gamma-ray
  background}},  \href{http://xxx.lanl.gov/abs/1202.5309}{{\tt
  arXiv:1202.5309}}.

\bibitem{Inoue:2008pk}
Y.~Inoue and T.~Totani, {\it {The Blazar Sequence and the Cosmic Gamma-Ray
  Background Radiation in the Fermi Era}},  {\em Astrophys. J.} {\bf 702}
  (2009) 523--536, [\href{http://xxx.lanl.gov/abs/0810.3580}{{\tt
  arXiv:0810.3580}}].

\bibitem{Abdo:2010nz}
{\bf Fermi-LAT Collaboration} Collaboration, A.~A. Abdo {\em et.~al.}, {\it
  {The Spectrum of the Isotropic Diffuse Gamma-Ray Emission Derived From
  First-Year Fermi Large Area Telescope Data}},  {\em Phys. Rev. Lett.} {\bf
  104} (2010) 101101, [\href{http://xxx.lanl.gov/abs/1002.3603}{{\tt
  arXiv:1002.3603}}].

\bibitem{Inoue10a}
Y.~Inoue {\em et.~al.}, {\it {The Cosmological Evolution of Blazars and the
  Extragalactic Gamma-Ray Background in the Fermi Era}},
  \href{http://xxx.lanl.gov/abs/1001.0103}{{\tt arXiv:1001.0103}}.

\bibitem{Xia:2011ax}
J.-Q. Xia, A.~Cuoco, E.~Branchini, M.~Fornasa, and M.~Viel, {\it {A
  cross-correlation study of the Fermi-LAT $\gamma$-ray diffuse extragalactic
  signal}},  {\em Mon.Not.Roy.Astron.Soc.} {\bf 416} (2011) 2247--2264,
  [\href{http://xxx.lanl.gov/abs/1103.4861}{{\tt arXiv:1103.4861}}].

\bibitem{Ackermann:2012uf}
M.~Ackermann, M.~Ajello, A.~Albert, L.~Baldini, J.~Ballet, {\em et.~al.}, {\it
  {Anisotropies in the diffuse gamma-ray background measured by the Fermi
  LAT}},  \href{http://xxx.lanl.gov/abs/1202.2856}{{\tt arXiv:1202.2856}}.

\bibitem{Komatsu:2010fb}
{\bf WMAP Collaboration} Collaboration, E.~Komatsu {\em et.~al.}, {\it
  {Seven-Year Wilkinson Microwave Anisotropy Probe (WMAP) Observations:
  Cosmological Interpretation}},  {\em Astrophys.J.Suppl.} {\bf 192} (2011) 18,
  [\href{http://xxx.lanl.gov/abs/1001.4538}{{\tt arXiv:1001.4538}}].

\bibitem{Ajello:2011zi}
M.~Ajello, M.~Shaw, R.~Romani, C.~Dermer, L.~Costamante, {\em et.~al.}, {\it
  {The Luminosity Function of Fermi-detected Flat-Spectrum Radio Quasars}},
  {\em Astrophys.J.} {\bf 751} (2012) 108,
  [\href{http://xxx.lanl.gov/abs/1110.3787}{{\tt arXiv:1110.3787}}].

\bibitem{Abdo:2010ge}
{\bf Fermi-LAT Collaboration} Collaboration, A.~A. Abdo {\em et.~al.}, {\it
  {The First Catalog of Active Galactic Nuclei Detected by the Fermi Large Area
  Telescope}},  {\em Astrophys. J.} {\bf 715} (2010) 429--457,
  [\href{http://xxx.lanl.gov/abs/1002.0150}{{\tt arXiv:1002.0150}}].

\bibitem{Falcke:2003ia}
H.~Falcke, E.~Koerding, and S.~Markoff, {\it {A Scheme to Unify Low-Power
  Accreting Black Holes - Jet- Dominated Accretion Flows and the Radio/X-Ray
  Correlation}},  {\em Astron. Astrophys.} {\bf 414} (2004) 895--903,
  [\href{http://xxx.lanl.gov/abs/astro-ph/0305335}{{\tt astro-ph/0305335}}].

\bibitem{Merloni:2003aq}
A.~Merloni, S.~Heinz, and T.~Di~Matteo, {\it {A fundamental plane of black hole
  activity}},  {\em Mon. Not. Roy. Astron. Soc.} {\bf 345} (2003) 1057,
  [\href{http://xxx.lanl.gov/abs/astro-ph/0305261}{{\tt astro-ph/0305261}}].

\bibitem{Gallo:2005tf}
E.~Gallo {\em et.~al.}, {\it {A dark jet dominates the power output of the
  stellar black hole Cygnus X-1}},  {\em Nature} {\bf 436} (2005) 819--821,
  [\href{http://xxx.lanl.gov/abs/astro-ph/0508228}{{\tt astro-ph/0508228}}].

\bibitem{Marconi:2003tg}
A.~Marconi {\em et.~al.}, {\it {Local Supermassive Black Holes, Relics of
  Active Galactic Nuclei and the X-ray Background}},  {\em Mon. Not. Roy.
  Astron. Soc.} {\bf 351} (2004) 169,
  [\href{http://xxx.lanl.gov/abs/astro-ph/0311619}{{\tt astro-ph/0311619}}].

\bibitem{Ghisellini:2009rb}
G.~{Ghisellini}, L.~{Maraschi}, and F.~{Tavecchio}, {\it {The Fermi blazars'
  divide}},  {\em \mnras} {\bf 396} (June, 2009) L105--L109,
  [\href{http://xxx.lanl.gov/abs/0903.2043}{{\tt arXiv:0903.2043}}].

\bibitem{Abdo:2009iq}
{\bf Fermi-LAT Collaboration} Collaboration, A.~A. Abdo {\em et.~al.}, {\it
  {The Spectral Energy Distribution of Fermi bright blazars}},  {\em Astrophys.
  J.} {\bf 716} (2010) 30--70, [\href{http://xxx.lanl.gov/abs/0912.2040}{{\tt
  arXiv:0912.2040}}].

\bibitem{Collaboration:2010yg}
T.~F.-L. Collaboration, {\it {Spectral Properties of Bright Fermi-detected
  Blazars in the Gamma-ray Band}},  {\em Astrophys.J.} {\bf 710} (2010)
  1271--1285, [\href{http://xxx.lanl.gov/abs/1001.4097}{{\tt
  arXiv:1001.4097}}].

\bibitem{Lott:2012tp}
B.~Lott, E.~Cavazzuti, S.~Cutini, D.~Gasparrini, and C.~Dermer, {\it {General
  Properties of Fermi/LAT Active Galactic Nuclei}},
  \href{http://xxx.lanl.gov/abs/1205.3038}{{\tt arXiv:1205.3038}}.

\bibitem{Gilmore:2009zb}
R.~C. Gilmore {\em et.~al.}, {\it {GeV Gamma-Ray Attenuation and the
  High-Redshift UV Background}},  {\em \mnras} {\bf 399} (Nov., 2009)
  1694--1708, [\href{http://xxx.lanl.gov/abs/0905.1144}{{\tt
  arXiv:0905.1144}}].

\bibitem{Abdo:2010kz}
{\bf Fermi-LAT Collaboration} Collaboration, A.~A. Abdo {\em et.~al.}, {\it
  {Fermi Large Area Telescope Constraints on the Gamma-ray Opacity of the
  Universe}},  {\em Astrophys. J.} {\bf 723} (2010) 1082--1096,
  [\href{http://xxx.lanl.gov/abs/1005.0996}{{\tt arXiv:1005.0996}}].

\bibitem{Ando:2006cr}
S.~Ando, E.~Komatsu, T.~Narumoto, and T.~Totani, {\it {Dark matter annihilation
  or unresolved astrophysical sources? Anisotropy probe of the origin of cosmic
  gamma-ray background}},  {\em Phys. Rev.} {\bf D75} (2007) 063519,
  [\href{http://xxx.lanl.gov/abs/astro-ph/0612467}{{\tt astro-ph/0612467}}].

\bibitem{Singal:2011yi}
J.~Singal, V.~Petrosian, and M.~Ajello, {\it {Flux and Photon Spectral Index
  Distributions of Fermi-LAT Blazars And Contribution To The Extragalactic
  Gamma-ray Background}},  {\em Astrophysical Journal} (2011)
  [\href{http://xxx.lanl.gov/abs/1106.3111}{{\tt arXiv:1106.3111}}].

\bibitem{Atwood:2009ez}
{\bf LAT} Collaboration, W.~B. Atwood {\em et.~al.}, {\it {The Large Area
  Telescope on the Fermi Gamma-ray Space Telescope Mission}},  {\em Astrophys.
  J.} {\bf 697} (2009) 1071--1102,
  [\href{http://xxx.lanl.gov/abs/0902.1089}{{\tt arXiv:0902.1089}}].

\bibitem{Pushkarev:2009dx}
A.~Pushkarev, Y.~Kovalev, M.~Lister, and T.~Savolainen, {\it {Jet opening
  angles and gamma-ray brightness of AGN}},  {\em \aap} {\bf 507} (Nov., 2009)
  L33--L36, [\href{http://xxx.lanl.gov/abs/0910.1813}{{\tt arXiv:0910.1813}}].

\bibitem{Meyer:2011uk}
E.~T. Meyer, G.~Fossati, M.~Georganopoulos, and M.~L. Lister, {\it {From the
  Blazar Sequence to the Blazar Envelope: Revisiting the Relativistic Jet
  Dichotomy in Radio-loud AGN}},  {\em Astrophys.J.} {\bf 740} (2011) 98,
  [\href{http://xxx.lanl.gov/abs/1107.5105}{{\tt arXiv:1107.5105}}].

\bibitem{Harding:2012sa}
J.~P. Harding and K.~N. Abazajian, {\it {Comment on 'Joint Anisotropy and
  Source Count Constraints on the Contribution of Blazars to the Diffuse
  Gamma-Ray Background'}},  \href{http://xxx.lanl.gov/abs/1204.3870}{{\tt
  arXiv:1204.3870}}.

\bibitem{Inoue08}
Y.~{Inoue}, T.~{Totani}, and Y.~{Ueda}, {\it {The Cosmic MeV Gamma-Ray
  Background and Hard X-Ray Spectra of Active Galactic Nuclei: Implications for
  the Origin of Hot AGN Coronae}},  {\em \apjl} {\bf 672} (Jan., 2008) L5--L8,
  [\href{http://xxx.lanl.gov/abs/0709.3877}{{\tt arXiv:0709.3877}}].

\bibitem{FaucherGiguere09}
C.~A. Faucher-Giguere and A.~Loeb, {\it {The Pulsar Contribution to the
  Gamma-Ray Background}},  {\em JCAP} {\bf 1001} (2010) 005,
  [\href{http://xxx.lanl.gov/abs/0904.3102}{{\tt arXiv:0904.3102}}].

\bibitem{Fields10}
B.~D. Fields, V.~Pavlidou, and T.~Prodanovic, {\it {Cosmic Gamma-Ray Background
  from Star-Forming Galaxies}},  {\em Astrophys.J.} {\bf 722} (2010) L199,
  [\href{http://xxx.lanl.gov/abs/1003.3647}{{\tt arXiv:1003.3647}}].

\bibitem{Chakraborty:2012sh}
N.~Chakraborty and B.~D. Fields, {\it {Inverse Compton Contribution to the
  Star-Forming Extragalactic Gamma-Ray Background}},
  \href{http://xxx.lanl.gov/abs/1206.0770}{{\tt arXiv:1206.0770}}.

\bibitem{Lacki:2012si}
B.~C. Lacki, S.~Horiuchi, and J.~F. Beacom, {\it {The Star-Forming Galaxy
  Contribution to the Cosmic MeV and GeV Gamma-Ray Background}},
  \href{http://xxx.lanl.gov/abs/1206.0772}{{\tt arXiv:1206.0772}}.

\bibitem{Ackermann:2012eba}
{\bf Fermi LAT Collaboration} Collaboration, M.~Ackermann {\em et.~al.}, {\it
  {GeV Observations of Star-forming Galaxies with \textit{Fermi} LAT}},  {\em
  Astrophys.J.} {\bf 755} (2012) 164,
  [\href{http://xxx.lanl.gov/abs/1206.1346}{{\tt arXiv:1206.1346}}].

\bibitem{Inoue:2011bm}
Y.~Inoue, {\it {Contribution of the Gamma-ray Loud Radio Galaxies Core
  Emissions to the Cosmic MeV and GeV Gamma-Ray Background Radiation}},  {\em
  Astrophys.J.} {\bf 733} (2011) 66,
  [\href{http://xxx.lanl.gov/abs/1103.3946}{{\tt arXiv:1103.3946}}].

\bibitem{Feldmann:2012rx}
R.~Feldmann, D.~Hooper, and N.~Y. Gnedin, {\it {Circum-Galactic Gas and the
  Isotropic Gamma Ray Background}},
  \href{http://xxx.lanl.gov/abs/1205.0249}{{\tt arXiv:1205.0249}}.

\bibitem{Ando:2005xg}
S.~Ando and E.~Komatsu, {\it {Anisotropy of the cosmic gamma-ray background
  from dark matter annihilation}},  {\em Phys. Rev.} {\bf D73} (2006) 023521,
  [\href{http://xxx.lanl.gov/abs/astro-ph/0512217}{{\tt astro-ph/0512217}}].

\bibitem{Miniati07}
F.~Miniati, S.~M. Koushiappas, and T.~Di~Matteo, {\it {Angular Anisotropies in
  the Cosmic Gamma-ray Background as a Probe of its Origin}},  {\em Astrophys.
  J.} {\bf 667} (2007) L1--L14,
  [\href{http://xxx.lanl.gov/abs/astro-ph/0702083}{{\tt astro-ph/0702083}}].

\bibitem{Baltz08}
E.~A. Baltz {\em et.~al.}, {\it {Pre-launch estimates for GLAST sensitivity to
  Dark Matter annihilation signals}},  {\em JCAP} {\bf 0807} (2008) 013,
  [\href{http://xxx.lanl.gov/abs/0806.2911}{{\tt arXiv:0806.2911}}].

\bibitem{SiegalGaskins:2008ge}
J.~M. Siegal-Gaskins, {\it {Revealing dark matter substructure with
  anisotropies in the diffuse gamma-ray background}},  {\em JCAP} {\bf 0810}
  (2008) 040, [\href{http://xxx.lanl.gov/abs/0807.1328}{{\tt
  arXiv:0807.1328}}].

\bibitem{SiegalGaskins:2009ux}
J.~M. Siegal-Gaskins and V.~Pavlidou, {\it {Robust identification of isotropic
  diffuse gamma rays from Galactic dark matter}},  {\em Phys. Rev. Lett.} {\bf
  102} (2009) 241301, [\href{http://xxx.lanl.gov/abs/0901.3776}{{\tt
  arXiv:0901.3776}}].

\bibitem{Hensley09}
B.~S. Hensley, J.~M. Siegal-Gaskins, and V.~Pavlidou, {\it {The detectability
  of dark matter annihilation with Fermi using the anisotropy energy spectrum
  of the gamma-ray background}},  {\em Astrophys. J.} {\bf 723} (2010)
  277--284, [\href{http://xxx.lanl.gov/abs/0912.1854}{{\tt arXiv:0912.1854}}].

\bibitem{Fornasa:2009qh}
M.~Fornasa, L.~Pieri, G.~Bertone, and E.~Branchini, {\it {Anisotropy probe of
  galactic and extra-galactic Dark Matter annihilations}},  {\em Phys. Rev.}
  {\bf D80} (2009) 023518, [\href{http://xxx.lanl.gov/abs/0901.2921}{{\tt
  arXiv:0901.2921}}].

\bibitem{Abazajian:2010zb}
K.~N. Abazajian, S.~Blanchet, and J.~Harding, {\it {Current and Future
  Constraints on Dark Matter from Prompt and Inverse-Compton Photon Emission in
  the Isotropic Diffuse Gamma-ray Background}},  {\em Phys.Rev.} {\bf D85}
  (2012) 043509, [\href{http://xxx.lanl.gov/abs/1011.5090}{{\tt
  arXiv:1011.5090}}].

\bibitem{Kneiske:2007jq}
T.~M. Kneiske and K.~Mannheim, {\it {BL Lac Contribution to the Extragalactic
  Gamma-Ray Background}},  {\em Astron.Astrophys.} {\bf 479} (Feb., 2007)
  41--47, [\href{http://xxx.lanl.gov/abs/0705.3778}{{\tt arXiv:0705.3778}}].

\bibitem{Venters:2010bq}
T.~M. Venters, {\it {Contribution to the Extragalactic Gamma-ray Background
  from the Cascades of Very-high Energy Gamma Rays from Blazars}},  {\em
  Astrophys.J.} {\bf 710} (2010) 1530--1540,
  [\href{http://xxx.lanl.gov/abs/1001.1363}{{\tt arXiv:1001.1363}}].

\bibitem{Venters:2012bx}
T.~M. Venters and V.~Pavlidou, {\it {Probing the Intergalactic Magnetic Field
  with the Anisotropy of the Extragalactic Gamma-ray Background}},
  \href{http://xxx.lanl.gov/abs/1201.4405}{{\tt arXiv:1201.4405}}.

\bibitem{Inoue:2012cs}
Y.~Inoue and K.~Ioka, {\it {Upper Limit on the Cosmological Gamma-ray
  Background}},  {\em Phys.Rev.} {\bf D86} (2012) 023003,
  [\href{http://xxx.lanl.gov/abs/1206.2923}{{\tt arXiv:1206.2923}}].

\end{thebibliography}\endgroup
\end{document}